\documentclass[12pt,aps,prd,superscriptaddress,amsfonts,amssymb,amsmath,byrevtex,showpacs,nofootinbib]{revtex4-1}
\usepackage[latin9]{inputenc}
\usepackage[english]{babel}
\usepackage{amsmath}
\usepackage{amssymb}
\usepackage{graphicx}
\usepackage{epstopdf}

\makeatletter
\usepackage{amsfonts}\usepackage{palatino}\usepackage{amsthm}\usepackage{epstopdf}
\usepackage{color}

\makeatother
\emergencystretch=\maxdimen
\begin{document}

\title{The Top-Down Complexity}

\author{Micha{\l } Mandrysz and Jakub Mielczarek \\
 Institute of Physics, Jagiellonian University, {\L }ojasiewicza
11, 30-348 Cracow, Poland }

\email{jakub.mielczarek@uj.edu.pl}
\email{michal.mandrysz@student.uj.edu.pl}

\begin{abstract}
The rising complexity of our terrestrial surrounding is an empirical fact. 
Details of this process evaded description in terms of physics for long time attracting attention and creating myriad of ideas including non-scientific ones.
In this essay we explain the phenomenon of the growth of complexity by combining 
our up to date understanding of cosmology, non-equilibrium physics and thermodynamics. 
We argue that the observed increase of complexity is causal in nature, stands in agreement with the second law of thermodynamics and has it's origin in the cosmological expansion.
Moreover, we highlight the connection between the leader of complexity growth in localized areas of space
with free energy rate density, starting from the largest scales towards the smaller ones. 
Finally, in the light of recent advances in non-equilibrium statistical 
mechanics, our belief in the causal structure of modern scientific theories 
is transferred to biological systems. 
On relevant scales, adaptation and complexity growth follows a similar pattern, in which free energy rate density is provided by an external electromagnetic radiation. 
The presented, holistic approach, arms us with predicting power about variety of attributes of complex systems and leads to a chain of successful explanations on all scales of the Universe. 

\end{abstract}

\maketitle
\selectlanguage{english}

The complexity of life on Earth can appear perplexing to our scientific apparatus. 
Simple microscopic laws that performed amazingly well in cases of small inanimate systems were in many cases less successful in the description of biological forms, let alone the ones with information processing capabilities (agency). This, however, does not mean that we need to discard this approach.
Rather, we realize that emergence involves \emph{history} and that we cannot understand microcosmos without comprehending the cosmos. 
Therefore, let us begin our discussion towards the origin of complexity rise on 
Earth from the largest scales perceived by the human mind. 

\begin{figure}[ht!]
\centering \includegraphics[width=17cm]{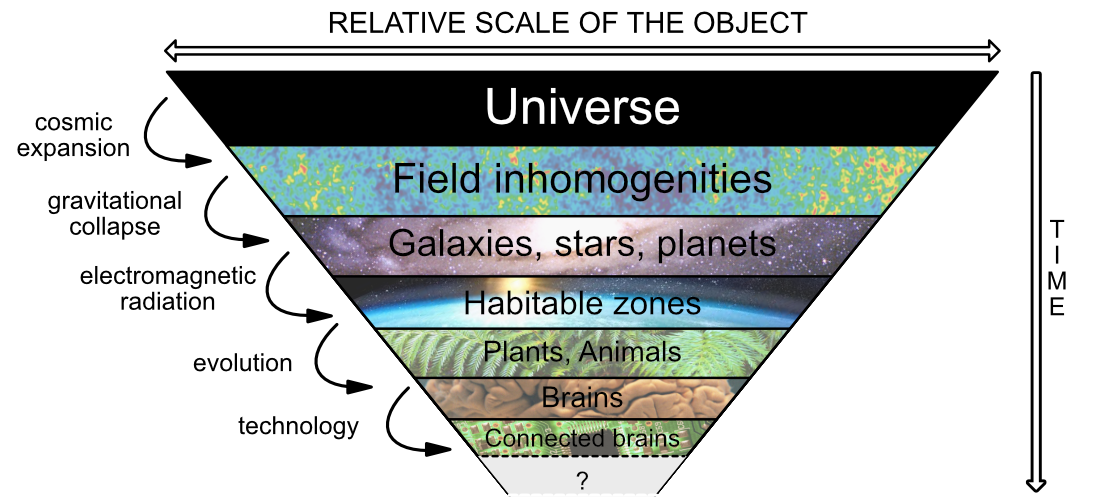} 
\caption{Top-Down trend of the complexity growth.}
\label{Fig1} 
\end{figure}

The Universe (See Fig. \ref{Fig1}), which we understand as basically everything that exist; 
is forming the largest, perfectly isolated system with no environment.

Therefore, if quantum mechanics applies, the Universe as a whole is 
described by a pure state $|\Psi\rangle$ satisfying the timeless  
Wheeler-DeWitt equation $\hat{H}|\Psi\rangle=0$, such that there
is no evolution of the state with respect to some external time 
parameter\footnote{This is just a reflection of the fact that there is no 
environment with respect to which the evolution of the system could 
be measured.}. However, \emph{internal times} are possible to introduce, 
which encode relative evolution between subsystems of the Universe. 

Since the state $|\Psi\rangle$ is pure it contains all information available 
in the universe. Consequently, for such a state with density matrix is $\rho=|\Psi\rangle\langle\Psi|$ the von Neumann 
entropy $S=-\text{tr}\left(\rho\ln\rho\right)$ remains constant. In the 
information theoretical picture, the von Neumann entropy can be 
considered as a measure of our lack of information
about the system. Therefore, for a fictitious observer, with an access to the whole 
state $|\Psi\rangle$ the total amount of information in the Universe might be considered 
a conserved quantity.

However, in general we deal with situations in which the observer can collect information only about a limited number of degrees of freedom. Such subset of the Universe we call the \emph{system}. The rest we call the 
\emph{environment}.

In the cosmological context, the Universe can be decomposed into background (the system) and 
inhomogeneities (the environment). 
As the studies for perturbative inhomogeneities suggest, in such 
a case the value of entanglement entropy and the value of the scale factor $a$ are positively correlated \cite{Zeh:2007uw, Kiefer:2017iy}. 
Consequently, in the present phase of the Universe expansion the entanglement 
entropy $S$ is increasing, satisfying the second law of thermodynamics. This, however, does not mean that the Universe will become homogeneous in the whole space, on the contrary islands of complexity are indeed possible as will shortly become clear.

\begin{figure}[ht!]
\centering \includegraphics[width=18cm]{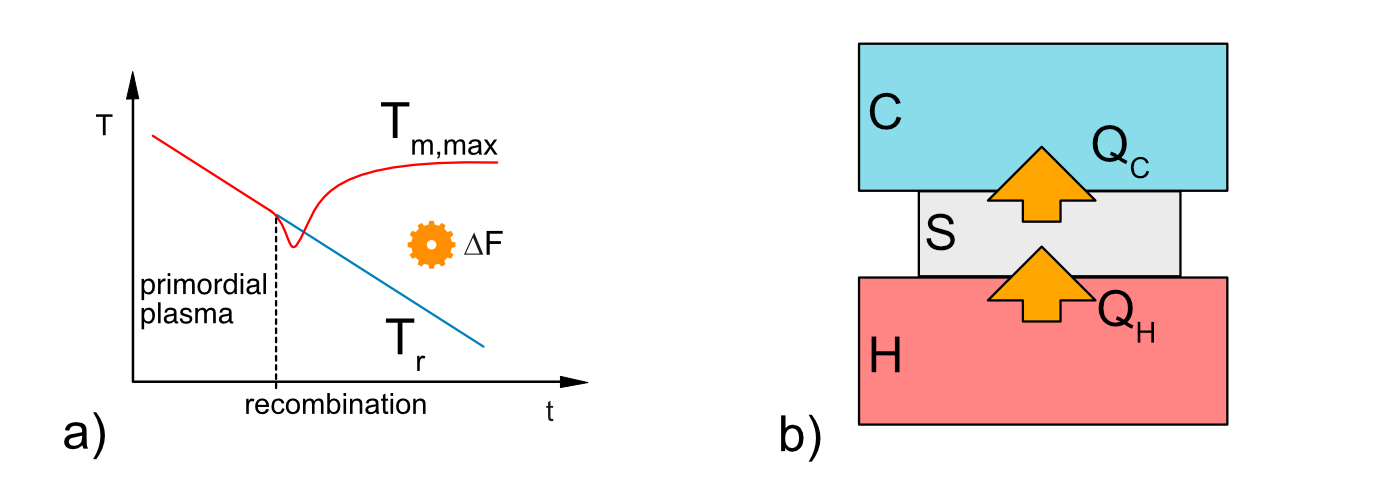} \caption{a) Thermal 
decoupling of radiation and matter. b) Schema of a heat engine consisting of the system (S) and its environment in thermal 
equilibrium (Heater $H$, Cooler $C$).}
\label{Fig2} 
\end{figure}

At the end of the radiation epoch the inhomogeneity of the Universe was
at the level $\Delta T/T \sim 10^{-5}$ and baryonic matter and radiation 
remained in the thermal equilibrium forming a primordial plasma (see Fig. \ref{Fig2}a). Then, 
as a consequence of the temperature drop, the recombination process 
(formation of neutral matter) occurred leading to departure of the system 
from the thermal equilibrium. This process can be perceived as a decay 
of the system into two equilibrium states each characterized by a different 
temperature (thermal decoupling). For radiation, temperature gradually 
decreased following the $T_r\sim1/a$ trend, which was valid also for 
the relativistic plasma before the recombination. 

The cosmic expansion enhanced inhomogeneities of the matter density, 
which ultimately became unstable initiating the process of gravitational collapse. 
This process lead to formation of dense inhomogeneous structures (stars) in which gravitational 
potential energy was transformed into the kinetic energy, increasing their temperatures.
Quite often the details of this process (concerning both the formation of stars and planets) are neglected and lead to confusion whether matter localized as stars or planets represents a state of higher entropy.
Therefore, let us articulate this clearly - in case of non-interacting particles the state of maximum entropy is a homogenous distribution. However, in case of gravitationally interacting particles the entropic ``price" for localizing particles in one place is ``paid" by the heat emitted to the environment, such that the total entropy in this process also rises. 
For stars the accumulated energy due to this process leads to ignition of 
the nuclear reactions increasing further the temperature of stars. 

The maximal temperature of localized matter systems (stars) $T_{m,max}$ 
eventually became much higher\footnote{Nevertheless, the 
averaged matter temperature remained below the temperature of radiation.} than the temperature of the background 
radiation $T_r$ filling the Universe (See Fig. \ref{Fig2}a).

This difference of temperatures between the hot spots (stars) and cold 
surrounding allows for work or the free energy $\Delta F$ (depicted as cogs in the illustrations) to be extracted as it 
happens in case of a heat engine (See Fig. \ref{Fig2}b). The stars are the heat reservoirs 
here while the cold space plays the role of the cooler. Objects existing close to the stars can obviously benefit by absorbing the incoming radiation and transforming it into usable work. These 
objects are called planets. 

The planets or rather planetary atmospheres are open systems resembling heat engines powered by solar radiation. Simple calculation yields that incoming solar photon may give rise to about 20 outgoing terrestrial 
infrared photons. Each photon has approximately the same entropy of order 
- one bit, so the entropy carried by the photons increases 20-fold which is more 
than enough to allow some decrease of the entropy of the life forms. The 
total entropy will go up even if the organisms manage to reduce their disorder. 
This of course only settles the consistence of the second law of thermodynamics, 
but does little to answer the question of why we observe the emergence of 
complex structures. This is why we will now perform an analysis of a simple non-equilibrium model that resembles 
the systems described in previous paragraphs and demonstrate that a system 
within that model lowers it's entropy.

Consider a system $S$, stacked in between a heater $H$ and a cooler $C$, 
much bigger than the system and with constant temperatures 
$T_H$ and $T_C$ (See Fig. \ref{Fig2}b) such that a steady amount of heat flows from the heater 
through the system to the cooler.

Following Prigogine's approach \cite{Prigogine:1978kz} we will analyze it from the 
perspective of internal entropy produced $(i)$ and external entropy transferred $(e)$ 
\emph{to} the system of interest $S$. Of course the entropy change of the system $dS_S$ 
would be the sum of those two contributions:
\begin{center}
$\frac{dS_S}{dt}=\frac{dS_i}{dt}+\frac{dS_e}{dt}$.
\end{center}
As we have mentioned, in our analysis we limit ourselves to the scenario of a \emph{stable 
state} in which the same amount of heat that goes in also goes out. In other words
$dQ_C=-dQ_H$, using which we get the following equation
\begin{center}
$dS_e=\frac{dQ_H}{T_H}+\frac{dQ_C}{T_C}=dQ_H\left(\frac{1}{T_H}-\frac{1}{T_C}\right)
=dQ_H\left(\frac{T_C-T_H}{T_HT_C}\right)<0$.
\end{center}
Therefore, the heat flux transfers some of the entropy outside the system (or negative 
entropy in). Now, since we require the state to be stable and the system's entropy can't 
grow ad infinitum we have \(dS_S=0\) and thus the external negative entropy flow 
balances the internal entropy production $dS_i=-dS_e>0$ (In some sense we could say 
that the rate of internal entropy production $dS_i$ is a function of the entropy inflow $dS_e$).

Our system is peculiar in one simple manner, namely, thanks to the constant heat flow, 
the system entropy decreases\footnote{To see this, Taylor expand $\frac{dS_i}{dt}=
j_i(S_S)=j_i\left(S_0\right)+\left(S_S-S_0\right)C_1+\mathcal{O}\left(S_S^2\right)$ 
and solve differential equation $\frac{dS_S}{dt}=j_e + j_i\left(S_S\right)=j_e +\frac{S_0-S_S}{\tau}$, 
getting $S_S(t)=S_0+j_e\tau \left(1-e^{-t/\tau }\right)$, where $j_e$ is a negative constant. 
In the $t\rightarrow \infty$  limit the system's entropy falls from the initial value 
$S(t=0)=S_0$ to the minimal value $S_{min}=S(t\rightarrow \infty) =S_0+j_e \tau < S_0$.}!
Non-equilibrium conditions indeed let us escape the tyranny of the entropy. 

When we turn on the driving force (here temperature difference) the entropy smoothly lowers to 
minimum entropy $S_{min}$. Moreover, this model does not store energy, hence internal energy 
stays constant, which means that the free energy, defined by $F~=~U~-~T~S$ gets maximized. 
Free energy allows us to perform work which can be realized in different ways depending on the 
nature of the system. It can mean the formation of stars, emergence of complex behaviour in the 
atmosphere or circular convective motion in the Benard cell.

\begin{figure}[ht!]
\centering \includegraphics[width=18cm]{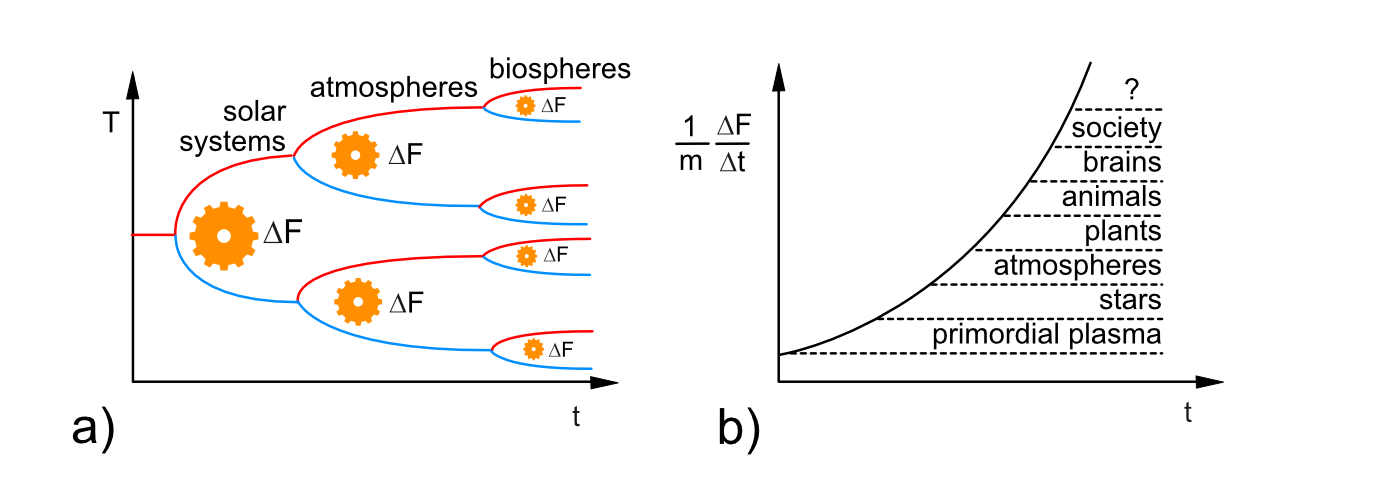} \caption{a) Emergence of heat engines 
powering complexity growth as a result of departure from equilibrium. b) Timeline of 
free energy rate density growth adopter from Ref. \cite{Chaisson:2001}.}
\label{Fig3} 
\end{figure}

In fact, Chaisson hypothesized \cite{Chaisson:2001} that for an object of mass $m$ the \emph{free energy rate density} 
$\frac{1}{m} \frac{\Delta F}{\Delta t}$ is a universal indicator of complexity of the system and 
that cosmic evolution is correlated with the growth of localized free energy rate density.
Indeed, this trend is visible from our top-bottom approach in which cosmic expansion leads 
to thermal decoupling (formation of a heat engine). Then, going to smaller and smaller scales this free energy is lent and localized in the subsystems resembling a bifurcation phenomena (See Fig. \ref{Fig3}a). First the free energy is passed from the solar systems to planetary atmospheres and then from planetary atmospheres to the biospheres on human scales. This trend does not seem to stop here and the free energy continues to get localized in human societies.

At first it may appear counterintuitive, but the free energy rate density is indeed smaller 
for stars than it is for planetary atmospheres, and it is smaller for planetary atmospheres than it is for animals.
The XXI century inventions, like engines and computers also have their place in it and might be considered a part of our modernised society, amounting to our complexity.
Human brains take the top places in this ladder of complexity and continuously climb it; through augmentation with technology.
 
We did not prove the association made by Chaisson by any rigorous way. At the end some further 
improvements might be possible, but the trend is clearly visible, almost as in Darwinian 
evolution. The inflows of free energy localize on smaller and smaller scales in tandem with growth of complexity.  

On the other hand, one shall not be tempted to try to find the explanation in 
terms of teleology, a way which has a certain appeal for human beings.
A great example can be given by trying to explain the convection currents in Benard cell. 
The sentence ``Benard cell activates another mean of transport to efficiently transport 
heat from the heater to the cooler" implies an idea proposed by the Maximum Entropy Production 
(MEP) principle which validity is under recent theoretical investigations \cite{Ozawa:2003jt}. 
Precisely because the MEP principle is currently teleological in nature it does not get a very wide approval 
among scientific community. Laymen might think that it's some close-minded thinking of 
the scientific community, but this insistence on causal explanations is one of main reasons 
of triumph and an enormous success of science in explaining natural phenomena. 
Advancement in our understanding of say, gravity, was made by the retreat from teleological explanations. 
For example the ideas of ancient greeks who believed that the cause of the downward motion of heavy bodies on 
Earth was related to their nature, which caused them to move downward, toward the center 
of the Universe, which was their natural place. Conversely, light bodies such as the element 
fire, moved by their nature upward toward the inner surface of the sphere of the Moon 
\cite{Pedersen:99571,Grant:2009ue}. 

Nowadays we recognize the naivety of those explanations and the superiority of Newtonian gravity. 
Even though the benefits of adopting Newtonian ideas were enormous as they allowed us to predict various events, they still left some room for improvements (force had instantaneous effect on distant bodies) as Newton himself noted \cite{Janiak:2006uw}.
This and more weaknesses were brilliantly noticed by Einstein, whose theory not only opened new areas of 
inquiry (black holes and studies of the beginning of the Universe) and made gravitational predictions
more precise (allowing us to develop GPS systems etc.), but also strengthened the \emph{causal} nature of gravity.

It's not as to say that teleology can not be a vestibule to physics. If the idea is further developed, 
a teleological explanation may be untangled to ``theory of principle" or a ``constructive theory", 
classification proposed by Einstein in 1919 \cite{Einstein:104771}. With enough work, MEP 
principle (in one form or another) might become a part of a valid scientific framework leading us to a correct and complete description of non-equilibrium phenomena.

So far we have learned that teleological explanations have little place in physics, but what might be
less obvious is that it is also the case for evolutionary biology. To see this, let's try to discriminate the 
content of various statements. The everyday statement, we usually consider true ``birds have 
wings \emph{to} fly'' has little scientific relevance. A slightly modified version of this statement ``birds 
evolved wings to fly'' is much better, because (even though it contains the grammatical particle ``to") it underlines it's causal ingredient, namely evolution. Natural selection is all about discovering the evolutionary pressures that led to 
formation of complex features; there is no goal in this process. Of course we talk about ``survival 
of the fittest", but this is just a poetic way of the description and the ultimate end state (goal) is not known. 
This feature of biology convinces us that it can, and will be ultimately described by physics and chemistry. 

To make the case stronger, just recently J. England \cite{England:2015hl} proposed a physical process by 
which evolution can take place at the microscopic scale. In fact his approach is another 
facade of the non-equilibrium theory, taste of which we experienced in the previous paragraphs.
In his paper, J. England employs a peculiar family of laws known as Fluctuation Theorems. 
In contrast to known laws of Statistical Mechanics those laws take into account the history 
of the process and discriminate between the more likely histories according to the entropy 
transferred to the environment. The more entropy (or in other words heat) is produced the 
more likely a microscopic path ($\Rightarrow$) is relative to it's reverse ($\Leftarrow$):
\begin{center}
$\frac{\Rightarrow}{\Leftarrow}=\exp\Delta S_{env}$. 
\end{center}
Once again, the end states are determined by microscopic evolution, the end goal is not set. 
Moreover, this last equation can be easily expanded to macrostates which give us a generalized form of second law of thermodynamics
\begin{center}
$\Delta S_{tot}+\ln(\frac{\Leftarrow}{\Rightarrow})>0$,
\end{center} (note the
change of direction of the arrows) where $\Delta S_{tot}$ is the
sum of entropy change of the environment and the system under consideration.

What this tells us is that the more irreversible a transition is, the less entropy has to be 
produced \emph{inside} the system (in a similar way to our previous considerations). 
However, the form of this equation makes the link between irreversibility and entropy 
much more evident and fruitful, thus we will mention here a few results to which it leads. 

If we apply it to a system under periodic driving force, then it was shown \cite{Perunov:2016hl} how the system can transform or remodel to a form in which it dissipates energy more efficiently. This result gave a pioneer example of evolution taking place at the microscale.
On the other hand applying this equation to self-replicators \cite{doi:10.1063/1.4818538} an inverse relation between the maximal growth rate of a self-replicator and it's durability was found.
This implies that there must be a biological tradeoff between the pace of reproduction and durability of self-replicators. 
Those advancements push us towards understanding the origin of biological complexity and it is likely that the next step, with the help of information theory, will unravel the emergence of agency.
In fact, purpose directed behaviour could be seen as a result of information processing and storage which in itself relies on availability of free energy as shown by Landauer \cite{Landauer:2002wc}.
This accumulation of knowledge, being synonymous to the \emph{free energy rate density} growth is also expected to follow the double exponent rule \cite{Kurzweil:2007vy}. 
Those and other insights guide as in understanding of complex phenomena and allow us to extrapolate the trends and prognose the future.

To sum up, we have presented a raw mechanism, by which localized complexity 
(stars, atmosphere, life, technology) emerges from physical laws of Nature. The 
crucial role was played by the free energy rate density, given by the chain of spontaneously 
forming heat engines initiated by cosmic expansion. We then argued that teleological 
reasoning can also be thought as an imperative result of the growing complexity 
which at about XVI century has been superseded by modern, causal, scientific reasoning in 
a similar way in which walking on land superseded swimming in sea for our 
predecessors. Obviously we still use it (as we still enjoy swimming), but it's purpose 
is different. Finally, thanks to recent advancement in non-equilibrium statistical 
mechanics, our belief in the causal structure and the illuminating power of modern scientific theories 
was transferred to the smaller scales governed by biology. There, adaptation took 
place under the influence of an external periodic force, in case of Earth - solar radiation.

The upcoming years will certainly bring us closer towards understanding the emergence of life and agency, in the context of causal frameworks of physics. Therefore, the insistence on finding purpose in the laws of Nature is illusionary and futile. 
This, however, should not be perceived negatively or discourage us, as we believe that through ingenuity and the pursuit of truth we will find ourselves a purpose.

\medskip
\medskip
\emph{``The significance of our lives and our fragile planet is then determined by our own wisdom and courage. We are the custodians of life's meaning. ...If we crave some cosmic purpose, then let us find ourselves a worthy goal." --- Carl Sagan, 1997 \cite{Sagan:1997up}}
\raggedbottom
\newpage

\bibliographystyle{ieeetr}
\bibliography{bibliography}

\end{document}